\documentstyle[draft%
,amscd%
,amssymb%
,12pt]{amsart}

\newtheorem{thm}{Theorem}[section]

\theoremstyle{definition}

\theoremstyle{remark}
\newtheorem{rem}{Remark}[section]

\numberwithin{equation}{section}

\newcommand{\shift}{\operatorname{sh}}

\begin{document}

\title[A note on simplicial dimension shifting]{
A note on simplicial dimension shifting}

\author{Adrian Ocneanu}

\address{Department of Mathematics, The Pennsylvania State University,
University Park, Pennsylvania 16801, USA}

\date{January 1993}

\maketitle

\begin{abstract}
We discuss a simplicial dimension shift which associates to each $n$-manifold
an $n-1$-manifold. As a
corollary we show that an invariant which was recently proposed by Ooguri and
by Crane and Yetter for
the construction of 4-dimensional quantum field theories out of 3-dimensional
theories is trivial.
\end{abstract}

Ooguri \cite{oog:top} proposed recently a method for obtaining 4-manifold
invariants and studying
quantum gravity, with a construction inspired statistical mechanics. This
procedure was further
formalized by Crane and Yetter \cite{crayet:top}. Their method is associating a
3-manifold with
boundary to every 4-simplex $\Delta^4$. Given 4-manifold $M$ with a
triangulation $\tau$ and an
invariant for 3-manifolds with boundary, they evaluate the invariant on the
3-manifolds with boundary
coming from each 4-simplex, they match boundary conditions according to the
triangulation and then
they compute a partition function. They mention that the proof of the
invariance of the result to the
choice of the triangulation is long and technical.

In fact the matching and summation of boundary conditions is equivalent to the
glueing together
of the corresponding portions of the boundary and then computing the invariant
of the 3-manifold
without boundary obtained this way. With this observation the choice of the
3-manifold invariant
becomes irrelevant. One obtains a topological construction, which associates to
each
triangulated 4-manifold a 3-manifold, which we call the dimension shift of $M$.
The aim of this
note is to prove that the dimension shift of a 4-manifold is essentially
trivial, and that
for a singular triangulated 4-manifold the dimension shift yields essentially
the direct sum of the
3-singularities of $M$. In particular, the invariants of Ooguri, Crane and
Yetter are always 1.

We were led independently to the same construction in a different context
\cite{ocn:top}. We showed an
equivalence between systems of n-modules over operator algebras and simplicial
invariants for
$n+1$-manifolds which have the glueing property. Working with subfactors, or
2-modules, an
asymptotic construction produces naturally some 3-modules. The 4-manifold
invariants associated to
these 3-modules are obtained from the 3-manifold invariants associated to the
2-modules by means of
dimension shifting.  Because of the result which follows we used this
construction only to show that
the axioms for 4-dimensional invariants are nonempty.

The idea of dimension shifting is the following. Let $\Delta^4$ denote the
4-simplex. The boundary
$\partial \Delta^4$ of $\Delta^4$ consists of 5 copies of the 3-simplex
$\Delta^3$, and is
topologically equivalent to the 3-sphere $S^3$.
% (one can picture the partition of $S^3$ e.g. as the 4
% tetrahedra obtained by dividing a tetrahedron $\Delta^3$ barycentrically,
%%together with the exterior
% $S^3 - \Delta^3$ as the 5-th simplex).
Consider the 1-skeleton $\sigma_1$ of $\partial \Delta^4$ and the dual
1-skeleton
$\widehat\sigma_1$ of $\partial \Delta^4$. Let $T$ be a
small tubular neighborhood of $\widehat\sigma_1$ in $S^3$.

Let now $M^4$ be a 4-manifold with a triangulation $\tau$. To each 4-simplex
$\Delta^4 \in \tau$
associate the 3-manifold with boundary $T$ as above. When two 3-faces
$\Delta^3$ and $\Delta'{}^3$ of
the 4-simplices $\Delta^4$ and $\Delta'{}^4$ in $\tau$ are glued, glue the
corresponding components
$\partial T\cap \Delta^3$ and  $\partial T'\cap \Delta'{}^3$ of the boundary of
$T$ and respectively
$T'$.

This way one obtains a 3-manifold without boundary denoted by
$\shift^{\operatorname{int}}_\tau M$.
In another version of this construction one glues the exterior $S^3 - T$ of $T$
the same way as
above, to obtain a 3-manifold $\shift^{\operatorname{ext}}_\tau M$. Note that
$S^3 - T$ is a
tubular neighborhood of the 1-skeleton $\sigma_1$.

\begin{thm} Let $M^4$ be a connected compact 4-manifold without boundary, with
a triangulation $\tau$
having $n_k$ simplices in each dimension $k=0,\dots,4$. Then we have
$$\shift^{\operatorname{int}}_\tau M = (10 n_4 - n_3 + 1) (S^1\times S^2)$$
and
$$\shift^{\operatorname{ext}}_\tau M = (n_1 - n_0 + 1) (S^1\times S^2).$$
\end{thm}

Here $m M$ denotes the connected sum of $m$ copies of the manifold $M$.

\begin{pf} The proof for the case of the interior dimension shift is the
following. Let $\Delta^4$ be
a 4-simplex in $\tau$, let $T$ be its thickened dual 1-skeleton and let
$\Delta^3$ be a 3-simplex in
$\partial \Delta^4$. Then $T\cap\Delta^3$ is topologically a 3-ball, and its
boundary consists of
$\partial T\cap\Delta^3$ and of $T\cap\partial\Delta^3$, which is the union of
4 copies of a 2-disk
$D^2$. When $\Delta^4$ is glued to another 4-simplex $\Delta'{}^4$ in $\tau$,
and $\Delta^3$ is
identified to its pair $\Delta'{}^3$ in $\partial \Delta'{}^4$, the balls
$T\cap \Delta^3$ and
$T'\cap \Delta'{}^3$ are glued together to yield a copy of $S^3$ which has a
3-ball hole $B^3$ carved
out for each face $\Delta^2$ of $\Delta^3$. The boundary $S^2$ of the hole
$B^3$ consists of a pair
of disks $D^2$ and $D'{}^2$ as described above. The disk $D^2$ is then glued to
a disk coming from
the neighbor $\Delta''{}^3$ of $\Delta^2$ in $\partial \Delta^4$ opposite to
$\Delta^3$.

This procedure yields a 3-sphere for every 3-simplex of $\tau$ and a handle
$S^2\times I$ for
every 2-face $\Delta^2$ of each 4-simplex $\Delta^4$ of $\tau$, for a total of
$10 n_4$ handles. From
these, $n_3 - 1$ handles are needed to connect the 3-spheres, and $10 n_4 - n_3
+ 1$ handles remain.

In the case of the exterior shift, note that the exterior $S^3-T$ corresponding
to a 4-simplex
$\Delta^4$ is a tubular neighborhood of the 1-skeleton of $\Delta^4$. We
decompose $S^3-T$ into
balls $B^3_v$ for each vertex $v$ of $\Delta^4$ joined by bars $b_e =
D^2_e\times I$, with $I$ the
1-simplex, obtained by thickening each 1-edge $e$ of $\Delta^4$.

Let now $v$ be a vertex of the triangulation $\tau$ of $M^4$. The simplicial
neighborhood, or star,
of $v$ is a cone over a 3-manifold $M_v$ called the link of $v$. The link $M_v$
is obtained by
joining the faces $\Delta^3_i$ opposed to $v$ in the 4-simplices $\Delta^4_i$
which contain
$v$. For each vertex $v$, each neighboring 4-simplex $\Delta^4_i$ contributes
to
$\shift^{\operatorname{ext}}_\tau M$ with a 3-ball $B^3_{vi}$.

The crucial observation is that the ball $B^3_{vi}$ is obtained by a homothety
from the 3-simplex
$\Delta^3_i$ opposite to $v$ in $\Delta^4_i$, and the glueing of the 3-balls
$B^3_{vi}$ is carried
over by this homothety from the glueing of the corresponding link simplices
$\Delta^3_i$. This
yields a 3-manifold $\widetilde M_v$ isomorphic to $M_v$ for each vertex $v$.
The manifold
$\widetilde M_v$ has a hole for each edge $e$ of $\tau$, with boundary obtained
by joining the disks
$D^2_{vei}$ which connect the balls $B_{vi}$ to the bars $b_{ei}$. By the same
homotopy, the disks
$D^2_{vei}$ join to make a 2-manifold $\widetilde M_e$ which is homothetic to
the link $M_e$ of
$e$. The bars themselves glue to give copies of $M_e\times I$.

For a manifold triangulation $\tau$ each $M_v$ is a 3-sphere and each $M_e$ is
a
2-sphere, and we obtain $n_0$ spheres $S^3$ and $n_1$ handles $S^2\times I$. A
counting argument
similar to the one done before ends the proof.
\end{pf} \begin{rem}

Analogous dimension shifts can be defined from an $n$-dimensional triangulated
manifold $M^n$ to an
$n-1$-manifold, using tubular neighborhoods of the $k$-skeleton of $\Delta^n$.
The proof above
extends to this context and shows that the interior dimension shift carries
essentially no
information at all about $M$. If $M$ is a triangulated manifold with
singularities, i.e.
such that the links of $M$ are not all spheres, then the exterior dimension
shift describes the
link singularities of $M$. Thus, for the 4-dimensional case described above,
the shifted
invariant of a singular triangulated 4-manifold is the product of the
3-dimensional
invariants of the links of its vertices. This is the result that we described
in \cite{ocn:top}.
\end{rem}

\bibliographystyle{amsplain}
\bibliography{dimensionShifting}

\end{document}